\documentclass[runningheads]{llncs}
\usepackage{graphicx}
\usepackage{multirow}
\usepackage{amsmath}
\usepackage{amssymb}
\usepackage{cite}
\usepackage[output-decimal-marker={,}]{siunitx}
\usepackage{enumitem}
\usepackage{epsfig}
\usepackage{booktabs}
\usepackage{tabularx}
\usepackage[outdir=./images/]{epstopdf}
\usepackage[caption=false,font=normalsize,labelfont=sf,textfont=sf]{subfig}
\begin{document}
\title{Speed Profile Definition for GLOSA Implementation on Buses Based on Statistical Analysis of Experimental Data
\thanks{This work is part of the project ”Joint Research Lab per la Mobilità urbana”. A particular acknowledgment goes to Azienda Trasporti Milanesi S.p.A. and Comune di Milano for the possibility of performing the road testing campaign, as well as  Vodafone Italia S.p.A. and Vodafone Automotive for the design and development of the communication architecture.}}% and the development of the SDK CITS-V2X for STEP, Safer Transport for Europe Platform, the Vodafone cooperative mobility solution that helps road users and road operators to make mobility safer, more secure and accessible.}}
\titlerunning{Speed Profile definition for GLOSA Implementation on Buses Based on Statistical Analysis}
\author{Daniele Vignarca \orcidID{0000-0003-2281-6038} \and
Stefano Arrigoni \orcidID{0000-0002-5316-7387} \and \\
Edoardo Sabbioni \orcidID{0000-0002-4356-8814} \and
Federico Cheli \orcidID{0000-0002-9807-5056}}
\authorrunning{D. Vignarca et al.}
\institute{Department of Mechanical Engineering, Politecnico di Milano, \\
Via La Masa 1, Milan 20156, Italy \\
\email{daniele.vignarca@polimi.it}}
\maketitle              % typeset the header of the contribution
\begin{abstract}
Intelligent Transportation Systems (ITS) are pushing an increasing interest and development when dealing with eco-driving systems. In this framework, this paper presents a method to define speed profiles specifically designed for Green Light Optimal Speed Advisory (GLOSA) systems on buses. GLOSA aims to optimize traffic flow by providing vehicles with real-time speed recommendations synchronized with traffic signal timings. Leveraging statistical analysis of experimental data collected from an urban bus, the study develops a methodology to extract meaningful insights into bus behaviour and traffic dynamics. The proposed approach considers road topology, scheduled bus stops, and signal timings to define simple although suitable speed profiles considering the peculiarities of the motion of a bus in an urban scenario. Through extensive data collection robust statistical data are defined, allowing the definition of vehicle motion profile for effectively develop and implement GLOSA systems. This research contributes to the advancement of Intelligent Transportation Systems by providing realistic data and practical insights for optimizing bus operations in urban environments.

\end{abstract}
%

%{\color{red} \bf The Extended Summary for AVEC’24 is 3 pages }

%{\color{red} \bf Please restrict your document to 6 pages for the Full paper of AVEC’24.}

%
\section{Introduction}
%Urban transportation systems are increasingly challenged to tackle environmental issues while simultaneously enhancing operational efficiency and passenger satisfaction. Among the array of strategies available, the optimization of bus operations has emerged as a promising avenue to meet these goals. 
The latest advancements in vehicle automation have revealed significant potential for enhancing traffic management via Advanced Driver Assist Systems (ADAS), benefiting both safety and environmental considerations. Green Light Optimal Speed Advisory (GLOSA) systems represent a significant application in the Cooperative-Intelligent Transportation System (C-ITS) field adopting Vehicle-to-Everything (V2X) communication technology \cite{Stahlmann2018}. Specifically, Green Light Optimal Speed Advisory (GLOSA) systems present an opportunity by providing real-time speed recommendations to vehicles, aiming to synchronize their movement with traffic light timings. This synchronization not only reduces fuel consumption, emissions, and travel time but also improves overall traffic flow \cite{Asadi2021, Simchon2020}. However, current GLOSA implementations predominantly target private vehicles, lacking the necessary customization to deal with urban bus peculiarities and its interaction with the infrastructure.

The literature nowadays addresses quite extensively the GLOSA, and C-ITS in general, for conventional vehicles like cars. At the same time, within this framework, there is emerging research starting to involve also the public transportation vehicles. The focus for buses is typically posed either on the comfort and regularity of the service for passengers \cite{Seredynski2014} or on the energy consumption which is reduced by adopting suitable speed profiles for the vehicle, thus avoiding unnecessary stops\cite{Shan2023}.
%The literature examining Green Light Optimal Speed Advisory (GLOSA) and Intelligent Transportation System (ITS) applications in urban transportation offers valuable insights into their potential benefits and challenges. 
\cite{Zhang2024} introduced an eco-driving strategy tailored for connected electric buses at signalized intersections, incorporating bus stops to underscore the advantages of multi-objective optimization in reducing energy consumption. In \cite{Ji2024} energy-saving profile planning for connected and automated electric buses is addressed, leveraging non-linear programming techniques to optimize speed profiles while considering motor characteristics. Beyond adjusting vehicle speed via GLOSA systems, alternative approaches like Transit Signal Priority (TSP) systems aim to extend traffic light green times and reduce red times during bus intersection negotiations \cite{Truong2019}. Additionally, the concept of Green Light Optimal Dwell Time Advisory (GLODTA) suggests increasing bus dwell time at stops \cite{Gallo2022}. Many studies advocate for a combined approach \cite{Zimmermann2021, Seredynski2020, Teng2023}, aiming to harness the comprehensive benefits of multiple systems.

In terms of modeling bus trajectories, \cite{Shan2023} distinguishes between the impact areas of bus stops and signalized intersections, ultimately formulating an optimization problem aimed at minimizing energy consumption through the use of the Dijkstra algorithm. The authors in \cite{Zhang2024} provide instead a comprehensive description of the bus stop negotiation process, taking into account the necessary time required for the vehicle to decelerate approaching the stop and to accelerate upon resuming its journey. All these studies showcase the increasing interest and progress in refining speed advisory systems for urban buses, highlighting the importance of customized strategies to enhance energy efficiency as well as passengers' comfort within public transportation networks.

This work presents a statistical analysis based on experimental data collected in real-world urban scenarios over one entire year. The outcome of this analysis allows the design of speed profiles typical for a public transportation vehicle, accounting for features such as the bus stop station for getting passengers off and on. The main contributions of the paper are the following:
\begin{itemize}
    \item Report real-world statistical data in terms of dwell time at different bus stops for different clusters based on the day of the week and time of the day.
    \item Define a set of suitable speed profiles for GLOSA implementation on buses (B-GLOSA), based on real vehicle parameters such as acceleration/deceleration as well as dwell time available from the statistical analysis.
\end{itemize}

The remainder of the paper is organized as follows: after the description in Section \ref{sec:exp_setup} of the experimental setup used for the data acquisition during the regular service of the bus, the statistical analysis of the collected data is presented in Section \ref{sec:stat_analysis}. Section \ref{sec:speed_prof} is then devoted to proposing different possible speed profiles for GLOSA implementation that can be designed based on the aforementioned statistical analysis. Finally, Section \ref{sec:conclusion} concludes the work and indicates possible future developments.

\section{Experimental data acquisition}\label{sec:exp_setup}
\begin{figure}[htbp]
\centering
\subfloat[][\emph{Sensors instrumentation}]{\includegraphics[width=.75\textwidth]{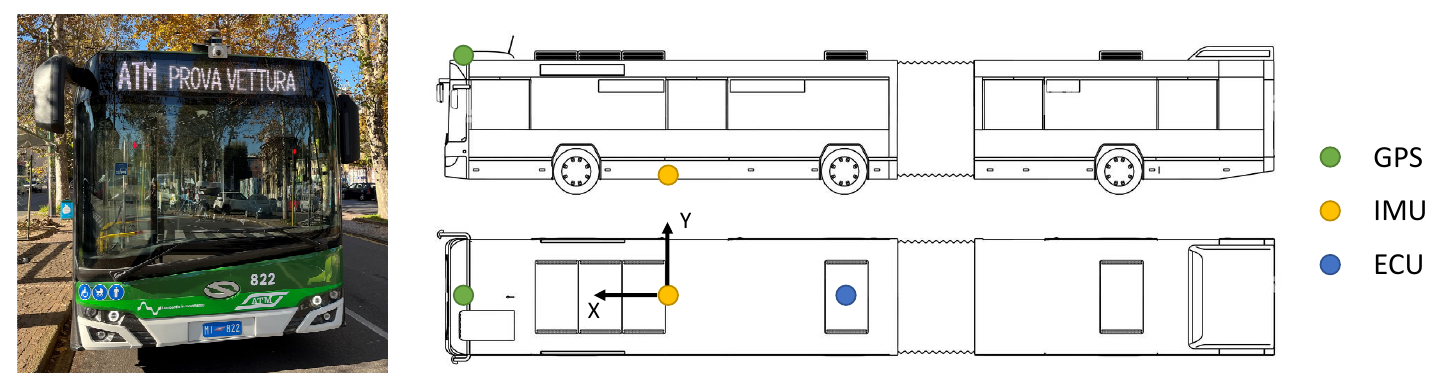}\label{fig:sensors}}
\subfloat[][\emph{Data collection}]{\includegraphics[width=.2\textwidth]{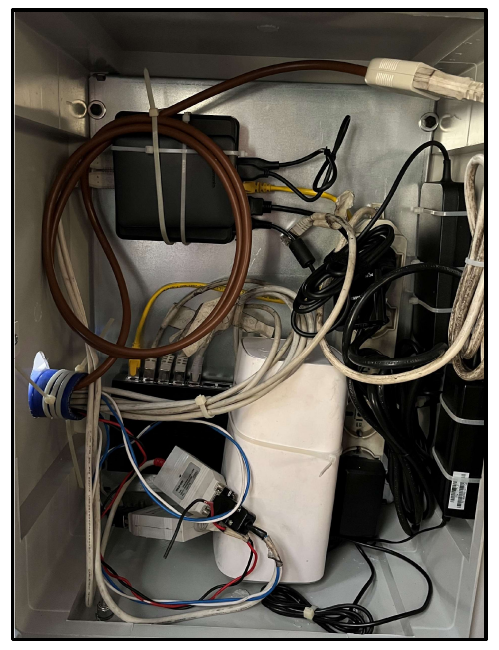}\label{fig:data_acq}} 
\caption{Schematics of the sensor and data acquisition setup on the vehicle.}
\label{fig:vehicle_scheme}
\end{figure}

This section presents the experimental setup for data acquisition during the vehicle's regular passenger service on the road. In particular, the vehicle shown in Fig. \ref{fig:sensors} is equipped with a Global Positioning System (GPS), an Inertia Measurement Unit (IMU), and the connection with the CAN-bus is established to read information coming from ECU such as vehicle speed and doors status. These data are used for vehicle localization presented in \cite{Vignarca2023mdpi}, as ego-vehicle position is then used to identify the bus stops present on the path.
As far as the data acquisition is concerned, the following elements have been enclosed into a compact architecture: (a) a computational unit (Intel NUC Core
i7 1165G7), operating on a soft real-time-based architecture using Robotic Operating System (ROS); (b) a 4 TB hard disk for data storage; (c) a 5G modem for internet connection; (d) a multi-port switch for the connection with sensors. All these devices, as well as the sensors, are powered by the 24V battery module for auxiliaries existing on the vehicle.

\section{Statistical Analysis}\label{sec:stat_analysis}
The present section is intended to report the results of the statistical analysis conducted on the basis of the data collected throughout the whole year. Indeed, the focus is on the dwell time the vehicle spends in correspondence with the 16 bus stops along the considered route (i.e., a portion of the route followed by the circular trolley-bus line 90/91 which goes around the city of Milan). 
Thanks to vehicle localization and the date-time reconstruction from recorded timestamps, it was possible to identify the bus stops along the path and cluster the dwelling times. It is worth mentioning on the one hand that the stop condition is detected by the combined matching of the following three conditions: (1) vehicle position located in the $\pm$ 20 meters around the bus stop position; (2) null vehicle speed and (3) opening of at least one of the four doors for passengers. On the other hand, samples associated with a stop duration shorter than 5 seconds and longer than 30 seconds are discarded, as they most likely are either not significant or affected by external factors.

\begin{figure}[htbp]
\centering
\subfloat[][\emph{Week-days}]{\includegraphics[width=.3\textwidth]{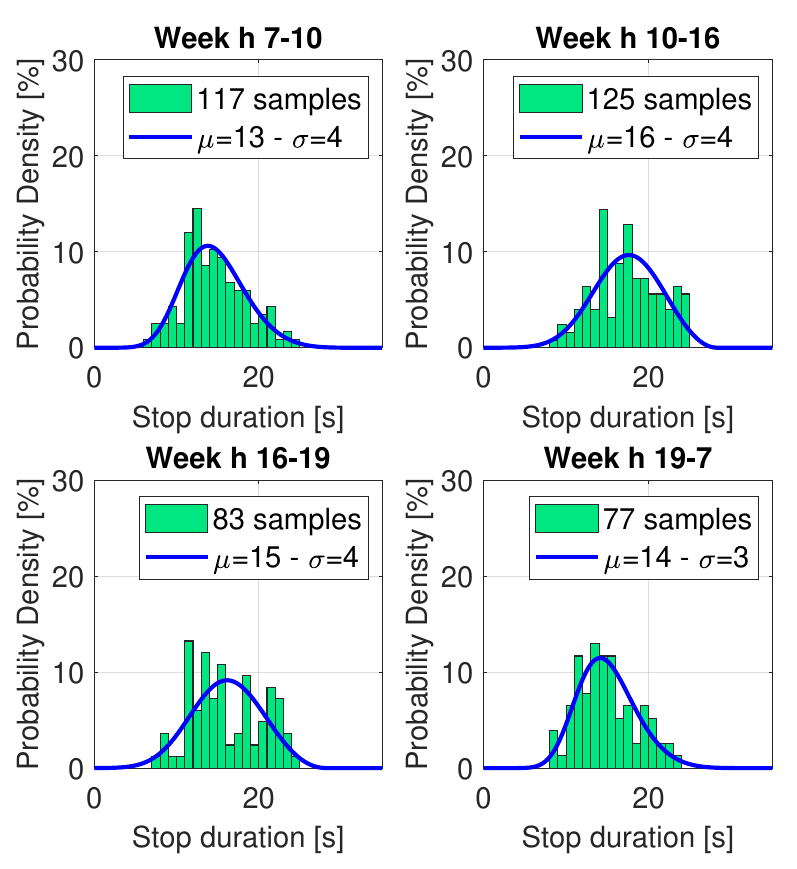}\label{fig:week}} \quad\quad
\subfloat[][\emph{Weekend-days}]{\includegraphics[width=.3\textwidth]{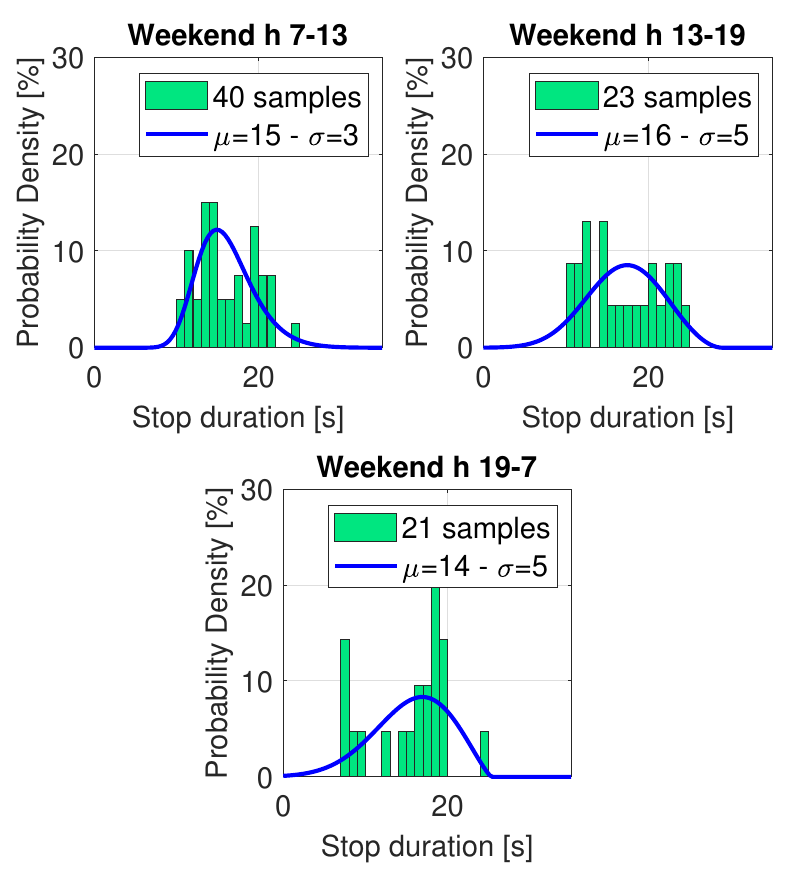}\label{fig:weekend}} 
\caption{GEV distribution fitting for different clusters at bus stop number 10 (i.e., central railway station).}
\label{fig:gevfit_plot}
\end{figure}

As mentioned, the data are categorized according to the day of the week and the time of the day. In particular, the weekdays are divided into the following four clusters: (1) \textbf{h 7-10}; (2) \textbf{h 10-16}; (3) \textbf{h 16-19}; (4) \textbf{h 19-7}. These clusters are associated with different peak and off-peak hours for workers and students. During the weekend days, the day hours are divided just for the morning (i.e., \textbf{h 7-13}) and the afternoon (i.e., \textbf{h 13-19}), keeping the same cluster for the night hours (i.e., \textbf{h 19-7}). Each cluster is then fitted using a Generalized Extreme Value (GEV) distribution, described by a mean value $\mu$ and a standard deviation $\sigma$. As an example, Fig. \ref{fig:gevfit_plot} reports the distributions for each cluster of one of the most crowded stops along the path (i.e., in correspondence of the central railway station). 

Table \ref{tab:stop_stat} summarizes the results of the measured dwelling time for the clusters mentioned above, for all the 16 bus stops considered, reporting the mean of the dwelling time, as well as its standard deviation. It is worth specifying that for some bus stops, especially when dealing with the weekend clusters, the data are not reported as the number of significant samples was limited or badly distributed, making the GEV fitting unreliable. As can be observed from the statistical data, some trends with the time of the day can be barely inferred. Indeed, the central hours of the day (i.e., \textbf{h 10-16}) are characterized by a higher flow of passengers, making the dwelling time longer on average. On the other hand, there seems to be a not-so-evident difference between week-days and weekend-days, as the route is part of a quite busy portion of the city with a considerable amount of passengers throughout the whole day.

\begin{table}[h!] 
\caption{Dwell time statistical data for different clusters.}
\centering
\begin{tabularx}{\textwidth}{ll ccc c ccc c ccc c ccc c ccc c ccc c ccc}
\toprule
&  & \multicolumn{16}{c}{Week}  & \multicolumn{11}{c}{Weekend} \\
&  & \multicolumn{3}{c}{h 7-10} & & \multicolumn{3}{c}{h 10-16} & & \multicolumn{3}{c}{h 16-19} & & \multicolumn{3}{c}{h 19-7} & & \multicolumn{3}{c}{h 7-13} & & \multicolumn{3}{c}{h 13-19} & & \multicolumn{3}{c}{h 19-7} \\
& \hspace{1mm} & $\mu$ [s] & $\sigma$ [s] & N & \hspace{2mm} & $\mu$ [s] & $\sigma$ [s] & N & \hspace{2mm} & $\mu$ [s] & $\sigma$ [s] & N & \hspace{2mm} & $\mu$ [s] & $\sigma$ [s] & N & \hspace{2mm} &  $\mu$ [s] & $\sigma$ [s] & N & \hspace{2mm} &  $\mu$ [s] & $\sigma$ [s] & N & \hspace{2mm} &  $\mu$ [s] & $\sigma$ [s] & N \\
\midrule
Stop 1  & & 12 & 4 & 97  & & 13 & 4 & 175  & & 14 & 3 & 121  & & 13 & 3 & 105  
        & & 14 & 4 & 61  & & 14 & 4 & 46   & & 13 & 4 & 28 \\
Stop 2  & & 11 & 4 & 110 & & 14 & 4 & 175  & & 14 & 4 & 108  & & 12 & 4 & 92 
        & & 12 & 4 & 59  & & 13 & 4 & 46   & & 13 & 4 & 29 \\
Stop 3  & & 14 & 4 & 92  & & 14 & 4 & 135  & & 13 & 3 & 98   & & 12 & 3 & 96   
        & & 12 & 3 & 57  & & 12 & 3 & 44   & & 13 & 4 & 29 \\
Stop 4  & & 12 & 4 & 103 & & 12 & 3 & 143  & & 12 & 3 & 106  & & 11 & 3 & 93    
        & & 11 & 3 & 49  & & 12 & 3 & 35   & & 12 & 3 & 28 \\
Stop 5  & & 14 & 5 & 95  & & 16 & 4 & 159  & & 16 & 4 & 104  & & 15 & 4 & 94    
        & & 17 & 4 & 50  & & -  & - & 34   & & 15 & 5 & 27 \\
Stop 6  & & 12 & 4 & 92  & & 12 & 4 & 122  & & 12 & 4 & 72   & & 11 & 3 & 66    
        & & 11 & 3 & 36  & & 14 & 4 & 40   & & -  & - & 9 \\
Stop 7  & & 14 & 4 & 119 & & 15 & 4 & 190  & & 13 & 3 & 117  & & 13 & 4 & 96    
        & & 15 & 4 & 42  & & 16 & 5 & 36   & & 14 & 4 & 21 \\
Stop 8  & & 11 & 3 & 114 & & 10 & 3 & 153  & & 10 & 3 & 106  & & 10 & 2 & 83    
        & & 9  & 2 & 38  & & 10 & 2 & 43   & & 11 & 3 & 15 \\
Stop 9  & & 12 & 3 & 98  & & 14 & 4 & 154  & & 12 & 4 & 103  & & 11 & 3 & 97    
        & & 11 & 3 & 46  & & 13 & 4 & 52   & & 11 & 3 & 26 \\
Stop 10 & & 13 & 4 & 117 & & 16 & 4 & 125  & & 15 & 4 & 83   & & 14 & 3 & 77    
        & & 15 & 3 & 40  & & 16 & 5 & 23   & & 14 & 5 & 21 \\
Stop 11 & & 11 & 3 & 104 & & 13 & 3 & 175  & & 12 & 4 & 102  & & 11 & 3 & 82    
        & & 13 & 4 & 46  & & 13 & 4 & 32   & & 13 & 3 & 25 \\
Stop 12 & & 15 & 4 & 94  & & 14 & 4 & 78   & & 13 & 3 & 73   & & 14 & 4 & 75    
        & & 13 & 3 & 44  & & 14 & 4 & 40   & & 12 & 2 & 22 \\
Stop 13 & & 11 & 3 & 99  & & 12 & 3 & 152  & & 11 & 3 & 113  & & 10 & 3 & 92    
        & & 1  & 3 & 47  & & 11 & 3 & 47   & & 10 & 2 & 17 \\
Stop 14 & & 13 & 5 & 100 & & 16 & 5 & 123  & & 16 & 5 & 92   & & 17 & 5 & 72   
        & & -  & - & 36  & & 14 & 6 & 35   & & -  & - & 16 \\
Stop 15 & & 11 & 3 & 108 & & 10 & 2 & 157  & & 10 & 2 & 111  & & 9  & 2 & 91    
        & & 9  & 2 & 43  & & 10 & 2 & 47   & & 10 & 2 & 24 \\
Stop 16 & & 12 & 3 & 108 & & 12 & 4 & 139  & & 11 & 3 & 107  & & 11 & 3 & 77     
        & & 11 & 3 & 48  & & 10 & 2 & 44   & & 12 & 4 & 20 \\
\midrule
Mean    & & 12 & 4 &     & & 13 & 4 &      & & 13 & 3 &      & & 12 & 3 &
        & & 12 & 3 &     & & 13 & 4 &      & & 12 & 3      \\
\bottomrule
\end{tabularx}
\label{tab:stop_stat}
\end{table}

\section{Speed profile definition}\label{sec:speed_prof}
As reported in the previous section, the information about the dwell time for every bus stop as a function of the date-time opens the possibility of designing speed profiles suitable for GLOSA implementation of public transport vehicles such as buses (B-GLOSA). In this case, the vehicle motion optimization has to consider that it is going to stop for sure at a given known point. Moreover, common GLOSA algorithms typically look for an optimal speed allowing the vehicle to cross the upcoming intersection with the green light is on. Within this process, the dwelling time of the bus is an additional variable to be taken into account when dealing with B-GLOSA design. As a result, having on this quantity a robust statistics to rely on, allows simplifying the modeling taking as a reasonable assumption a known constant dwelling time.

\begin{figure}[h]
\centerline{\includegraphics[width=0.5\textwidth]{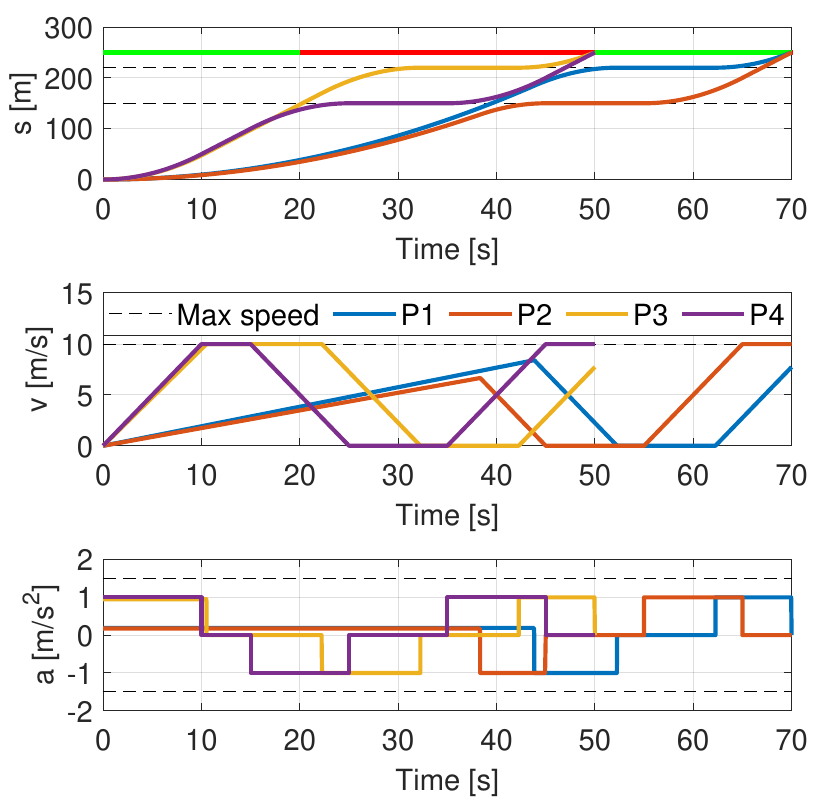}}
\caption{Designed possible speed profiles.}
\label{fig:profiles}
\end{figure}

\begin{figure}[h]
\centering
\subfloat[][\emph{Class choice}]{\includegraphics[width=.35\textwidth]{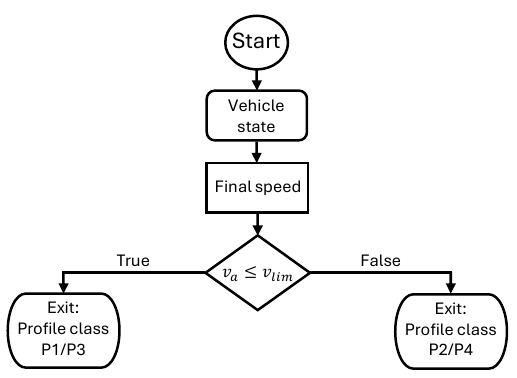}\label{fig:class_choice}} \quad \quad
\subfloat[][\emph{Time choice}]{\includegraphics[width=.4\textwidth]{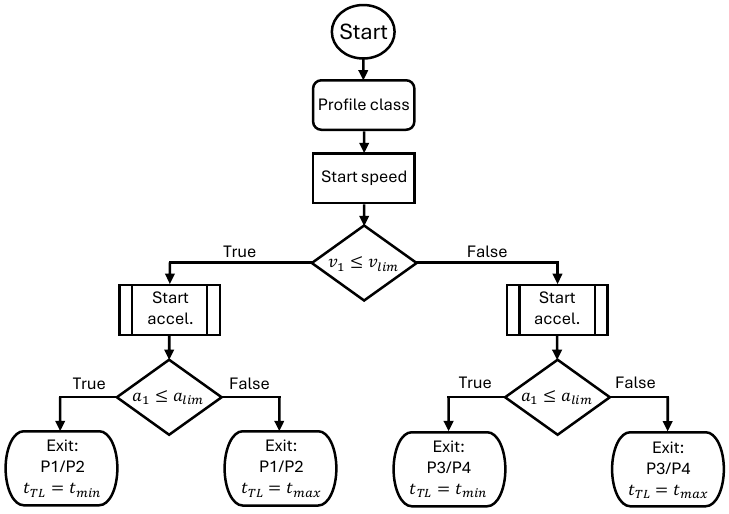}\label{fig:time_choice}} 
\caption{speed profile choice flowcharts.}
\label{fig:profile_choice}
\end{figure}

In this paper, four different possible speed profiles are proposed to address the B-GLOSA implementation. For the speed profile design, these quantities are assumed to be known: initial vehicle state (i.e., location along the path and speed), traffic light phases (i.e., minimum and maximum time available to reach the intersection with green light), bus stop location and corresponding dwell time (i.e., coming from statistics presented in Section \ref{sec:stat_analysis}), vehicle speed and acceleration limits (i.e., for both passengers safety and comfort issues). As far as the most suitable speed profile is concerned, it can be split into two sections: an initial one from the initial point to the bus stop and a final one from the bus stop up to the upcoming intersection. Each of these two sections is assumed to be either a simple uniformly accelerated motion or a uniformly accelerated motion followed by a constant speed motion. As a result, combining these options together, the four possible speed profiles shown in Fig. \ref{fig:profiles} are:
\begin{itemize}
    \item \textbf{P1}: uniform acceleration - brake - dwell - uniform acceleration;
    \item \textbf{P2}: uniform acceleration - brake - dwell - uniform acceleration - constant speed;
    \item \textbf{P3}: uniform acceleration - constant speed - brake - dwell - uniform acceleration;
    \item \textbf{P4}: uniform acceleration - constant speed - brake - dwell - uniform acceleration - constant speed.
\end{itemize}
It is worth highlighting that, since the acceleration for the second section of the profile (i.e., from stop to intersection) is fixed, the whole section is constrained by the geometric characteristic of the path (i.e., the distance between stop and traffic light). As a consequence, depending on this information the chosen profile will either have a constant speed part (i.e., P2/P4) or not (i.e., P1/P3) as described in the flowchart in Fig. \ref{fig:class_choice}, being the speed limit set to \SI{40}{km.h^{-1}} for safety reasons. 

Besides the topology of the route, the most affecting parameter is the time available to reach the intersection within the green light phase. In fact, to prioritize travel time reduction, the profile is designed so that the vehicle reaches the intersection at the switching time from red to green. The process followed for the timing choice is detailed in the flowchart reported in Fig. \ref{fig:time_choice}, being constrained the deceleration (i.e., \SI{-1}{m.s^{-2}}) and the acceleration of the final section (i.e., \SI{-1}{m.s^{-2}}). As a consequence of these assumptions, the remaining free variable for the speed profile design is the acceleration in the initial stage from the initial condition to the start of the braking manoeuvre.

\section{Conclusion}\label{sec:conclusion}
This research paper presented a robust statistical analysis based on real-world data collection throughout an entire year from an instrumented trolley bus during its regular passenger service. The outcome of these statistical results in terms of average dwelling time at the bus stop represented the basis for the speed profile definition when dealing with the implementation of a Green Light Optimal Speed Advisory system applied to local public transportation vehicles. In particular, four different speed profiles have been proposed leveraging topographic, vehicle performances, and timing constraints.
The availability of such a great amount of data from a bus real service allows to perform further and deeper analysis of the data collected on the road, thus representing a straightforward future development for this activity. On the other hand, the implementation and testing, both numerical and experimental, of a GLOSA system based on the proposed speed profiles would give the possibility to assess them from a quantitative point of view.


%\bibliographystyle{splncs04}
%\bibliography{AVEC2024_template/fulltext}

\begin{thebibliography}{10}
\providecommand{\url}[1]{\texttt{#1}}
\providecommand{\urlprefix}{URL }
\providecommand{\doi}[1]{https://doi.org/#1}

\bibitem{Asadi2021}
Asadi, M., Fathy, M., Mahini, H., Rahmani, A.M.: A systematic literature review of vehicle speed assistance in intelligent transportation system. IET Intelligent Transport Systems  \textbf{15},  973--986 (8 2021). \doi{10.1049/itr2.12077}

\bibitem{Gallo2022}
Gallo, F., Sacco, N.: On optimizing bus dwell times to reduce the probability of stopping for a red light at intersections. In: 2022 IEEE 25th International Conference on Intelligent Transportation Systems (ITSC). vol. 2022-October, pp. 1637--1644. Institute of Electrical and Electronics Engineers Inc. (2022). \doi{10.1109/ITSC55140.2022.9922295}

\bibitem{Ji2024}
Ji, J., Bie, Y., Shi, H., Wang, L.: Energy-saving speed profile planning for a connected and automated electric bus considering motor characteristic. Journal of Cleaner Production  \textbf{448} (4 2024). \doi{10.1016/j.jclepro.2024.141721}

\bibitem{Seredynski2020}
Seredynski, M., Laskaris, G., Viti, F.: Analysis of cooperative bus priority at traffic signals. IEEE Transactions on Intelligent Transportation Systems  \textbf{21},  1929--1940 (5 2020). \doi{10.1109/TITS.2019.2908521}

\bibitem{Seredynski2014}
Seredynski, M., Ruiz, P., Szczypiorski, K., Khadraoui, D.: Improving bus ride comfort using glosa-based dynamic speed optimisation. In: 2014 IEEE International Parallel \& Distributed Processing Symposium Workshops. pp. 457--463 (2014). \doi{10.1109/IPDPSW.2014.58}

\bibitem{Shan2023}
Shan, X., Wan, C., Hao, P., Wu, G., Zhang, X.: Connected eco-driving for electric buses along signalized arterials with bus stops. IET Intelligent Transport Systems  \textbf{17},  575--587 (3 2023). \doi{10.1049/itr2.12285}

\bibitem{Simchon2020}
Simchon, L., Rabinovici, R.: Real-time implementation of green light optimal speed advisory for electric vehicles. Vehicles  \textbf{2},  35--54 (3 2020). \doi{10.3390/vehicles2010003}

\bibitem{Stahlmann2018}
Stahlmann, R., Möller, M., Brauer, A., German, R., Eckhoff, D.: Exploring glosa systems in the field: Technical evaluation and results. Computer Communications  \textbf{120},  112--124 (5 2018). \doi{10.1016/j.comcom.2017.12.006}

\bibitem{Teng2023}
Teng, K., Liu, H., Liu, Q., Lu, X.: A cooperative control method combining signal control and speed control for transit with connected vehicle environment. IET Control Theory and Applications  (4 2023). \doi{10.1049/cth2.12608}

\bibitem{Truong2019}
Truong, L.T., Currie, G., Wallace, M., Gruyter, C.D., An, K.: Coordinated transit signal priority model considering stochastic bus arrival time. IEEE Transactions on Intelligent Transportation Systems  \textbf{20},  1269--1277 (4 2019). \doi{10.1109/TITS.2018.2844199}

\bibitem{Vignarca2023mdpi}
Vignarca, D., Arrigoni, S., Sabbioni, E.: Vehicle localization kalman filtering for traffic light advisor application in urban scenarios. Sensors  \textbf{23}(15) (2023). \doi{10.3390/s23156888}

\bibitem{Zhang2024}
Zhang, Y., Fu, R., Guo, Y., Yuan, W.: Eco-driving strategy for connected electric buses at the signalized intersection with a station. Transportation Research Part D: Transport and Environment  \textbf{128} (3 2024). \doi{10.1016/j.trd.2024.104076}

\bibitem{Zimmermann2021}
Zimmermann, L., Coelho, L.C., Kraus, W., Carlson, R.C., Koehler, L.A.: Bus trajectory optimization with holding, speed and traffic signal actuation in controlled transit systems. IEEE Access  \textbf{9},  143284--143294 (2021). \doi{10.1109/ACCESS.2021.3122087}

\end{thebibliography}

\end{document}